\documentclass[twocolumn,showpacs,reprint,amsmath,amssymb,eqsecnum,pre]{revtex4-1}

\usepackage{graphicx}      
\usepackage{dcolumn}       
\usepackage[usenames,dvipsnames]{color}

\newcommand{\sm}{\mathsf}
\newcommand{\cm}{\mathcal}
\newcommand{\beq}{\begin{equation}}
\newcommand{\eeq}{\end{equation}}
\newcommand{\beqn}{\begin{eqnarray}}
\newcommand{\eeqn}{\end{eqnarray}}

\newcommand\beqa{\begin{eqnarray}}
\newcommand\eeqa{\end{eqnarray}}
\newcommand{\nn}{\nonumber\\}
\newcommand{\dd}{\text{d}}

\newcommand{\NN}{\cm{N}}
\newcommand{\ii}{\text{i}}
\newcommand{\wt}{\widetilde}
\newcommand{\Q}{\mathcal{D}}
\newcommand{\dt}{D}

\definecolor{verde}{rgb}{.1,.4,0}
\definecolor{violet}{rgb}{.5,.0,1}
\definecolor{gray}{rgb}{.5,.5,.5}
\definecolor{brown}{rgb}{.4,.1,0}
\definecolor{orange}{rgb}{1,.5,0}
\definecolor{bordo}{rgb}{.5,0,.2}
\definecolor{rojo}{rgb}{.8,0,0}
\definecolor{otro}{rgb}{.9,0,0}

\newcommand{\ed}{\end{document}}

\begin{document}
\title{Exact solution of the Percus--Yevick integral equation for fluid mixtures of hard hyperspheres}

\author{Ren\'e D. Rohrmann}
\email{rohr@icate-conicet.gob.ar}
\homepage{http://icate-conicet.gob.ar/rohrmann}
\affiliation{Instituto de Ciencias Astron\'omicas, de la Tierra y del
Espacio (ICATE-CONICET), Avenida Espa\~na 1512 Sur, 5400 San Juan, Argentina}
\author{Andr\'es Santos}
\email{andres@unex.es}
\homepage{http://www.unex.es/eweb/fisteor/andres}
\affiliation{ Departamento de F\'{\i}sica, Universidad de Extremadura,
E-06071 Badajoz, Spain}

\date{\today}
\begin{abstract}
Structural and thermodynamic properties of
multicomponent hard-sphere fluids at odd dimensions have recently been derived in the
framework of the rational function approximation (RFA) [Rohrmann and Santos, Phys. Rev. E \textbf{83}, 011201 (2011)].
It is demonstrated here that the RFA technique yields the exact solution of
the Percus--Yevick (PY) closure to the Ornstein--Zernike (OZ) equation for binary
mixtures at arbitrary odd dimensions.
The proof relies mainly on the Fourier transforms $\widehat{c}_{ij}(k)$ of
the direct correlation functions defined by the OZ relation. From the analysis
of the poles of $\widehat{c}_{ij}(k)$ we show that the direct correlation
functions evaluated by the RFA method vanish outside the hard core, as
required by the PY theory.

\end{abstract}


\maketitle
\section{ Introduction} \label{s.intr}

The understanding and interpretation of thermodynamic properties and the structure
of real dense gases and liquids rely mostly on the use of models of fluids
and the application of approximate theories \cite{HM06}. One of the most successful and widely
used approximate methods is the Percus--Yevick (PY) theory. It is an integral
equation theory based on the Ornstein--Zernike (OZ) equation \cite{OZ14}
coupled with the PY closure \cite{PY58}. Systems of rigid
nonattracting hard spheres (HS), as well as their modified forms (e.g., square-well and sticky
particles), represent useful fluid models for the study of liquids and colloidal systems \cite{M08}.
In this context, it is noteworthy that exact solutions of the PY integral equation were early obtained for pure  \cite{W63,T63} and
multicomponent \cite{L64} HS fluids.

Recently, we have shown \cite{RS07,RS11} that the thermodynamic and
structural properties of single and (additive) multicomponent fluids of hard
hyperspheres at odd dimensions can be studied by means of an
analytical technique, the so-called rational function approximation (RFA).
This method is formulated in terms of the Laplace transform of a polynomial times the radial
distribution function  and leads directly to a system of algebraic equations
that have analytical solution at low spatial dimensionality (odd $d\le 7$ for
pure fluids and odd $d\le 3$ for mixtures) and must be solved numerically at
higher odd dimensions.

In the single-component case, it was shown that, in its simplest formulation, the RFA method recovers the known
solution of the PY closure to the OZ equation \cite{RS07}.
In the multicomponent case, the equivalence between the solutions of the two
approximations has been shown for binary mixtures in three dimensions \cite{YSH98}
and five dimensions \cite{RS11}. Although compelling arguments were presented in Ref.\ \cite{RS11} in favor of the extension of the equivalence to general odd $d$, a rigorous proof was not attempted.
Therefore, whether or not the PY integral equation is exactly solvable for mixtures of additive hard hyperspheres with arbitrary $d=\text{odd}$  (as well as the structure of the solution in the affirmative case) remained, to the best of our knowledge, an open problem.

The purpose of the present paper is to give a formal proof that the PY and RFA
solutions are indeed equivalent for binary mixtures in arbitrary odd dimensions.
The proof is based on the analysis of the Fourier transform $\widehat{c}_{ij}(k)$ of the direct
correlation function ${c}_{ij}(r)$, which is defined by the OZ equation
and  is a key quantity to be determined by both approximations.
The equivalence between RFA and PY theories is ensured by first deriving the functional structure of
$\widehat{c}_{ij}(k)$ as given by the RFA.
The direct correlation function ${c}_{ij}(r)$ is then calculated by the theory of
residues, showing that it is strictly short ranged, as expressed by the
PY closure.

The organization of the paper is as follows. Section \ref{s.formulation} provides a description of the fluid structure
equations and basic definitions of the two approximations.
Section \ref{s.binary} particularizes the RFA scheme to binary mixtures of HS
at odd dimensionalities.
Section \ref{s.analysis} deals with the analysis of the direct correlation
function and its short-range behavior as derived from the RFA theory.
We close the paper in Sec.\ \ref{s.conclusions} with some concluding remarks.

\section{Formulation} \label{s.formulation}

\subsection{Basic quantities}
Let us consider {an $\NN$-component}   fluid mixture of additive $d$-dimensional HS.
Let $\{x_i; i=1,\ldots,\NN\}$ be the set
of mole fractions,  $\{\sigma_i; i=1,\ldots,\NN\}$ be the set of diameters, and
$\rho$ be the total number density.

The radial distribution function $g_{ij}(r)$, the total correlation function
$h_{ij}(r)$, and the direct correlation function $c_{ij}(r)$ corresponding to
particle pairs of species $i$ and $j$ are the primary linkage between
thermodynamic and structural properties and particle interactions of multicomponent fluid mixtures.
They are not independent quantities. In particular $h_{ij}(r)=g_{ij}(r)-1$, while
$c_{ij}(r)$ is defined in terms of $h_{ij}(r)$ by the OZ
equation, which is conveniently written in  Fourier space and
in matrix form as
\beq \label{OZ}
\wt{\sm{c}}(k) = \sm{I} - \left[\sm{I}+\wt{\sm{h}}(k) \right]^{-1}=\wt{\sm{h}}(k)\cdot\left[\sm{I}+\wt{\sm{h}}(k) \right]^{-1}.
\eeq
Here, $\sm{I}$ is the $\NN\times\NN$ unit matrix and $\wt{\sm{c}}(k)$ and $\wt{\sm{h}}(k)$
are $\NN\times\NN$ matrices with elements $\rho\sqrt{x_ix_j}\widehat{c}_{ij}(k)$
and $\rho\sqrt{x_ix_j}\widehat{h}_{ij}(k)$, respectively, where the
hat symbol ($\,\widehat\,\,$)  denotes the Fourier transform.

The Fourier transform of the total correlation function  can be evaluated from
\beq \label{hG}
\widehat{h}_{ij}(k)=
\nu\left[\frac{G_{ij}(s)-G_{ij}(-s)}{s^{d-2}}\right]_{s=\ii k},\quad
\nu\equiv(-2\pi)^{(d-1)/2},
\eeq
where $\ii$ is the imaginary unit and  $G_{ij}(s)$ is a Laplace functional of the radial distribution function
defined by \cite{RS11}
\beq
G_{ij}(s)=\int_0^\infty \dd r\, r g_{ij}(r)\theta_n(sr)e^{-sr},
\label{Gij}
\eeq
with
\beq
\label{theta}
\theta_n(t)= \sum_{\ell=0}^{n} \omega_{n,\ell} t^\ell,\quad \omega_{n,\ell}\equiv
\frac{(2n-\ell)!}{2^{n-\ell}(n-\ell)!\ell!},
\eeq
being the reverse Bessel polynomial of degree $n=(d-3)/2$.
The function $G_{ij}(s)$ is a fundamental quantity in the sense that it
contains all the information about the thermodynamic properties of the fluid
and its knowledge gives structural information equivalent to pair radial
distributions, pair correlation functions, and structure factors.

Since the HS particles  are not allowed to penetrate  each
other, the radial distribution function vanishes inside the core, i.e.,
\beq \label{ghard}
g_{ij}(r)=1+h_{ij}(r)=0, \quad r<\sigma_{ij},
\eeq
where $\sigma_{ij}=\frac 12(\sigma_i+\sigma_j)$ is the contact distance for
hyperspheres of species $i$ and $j$.
In addition, the exact Laplace functionals $G_{ij}(s)$ for $d$-dimensional HS satisfy the following asymptotic relations \cite{RS11}.
\begin{itemize}
\item[(i)] Long wave-number limit:
\beq \label{Gsi}
\lim_{s\rightarrow \infty} s^{(5-d)/2}
e^{\sigma_{ij} s} G_{ij}(s)=\sigma_{ij}^{(d-1)/2}g_{ij}(\sigma_{ij}^+).
\eeq
\item[(ii)] Short wave-number expansion:
\beq \label{Gshort}
G_{ij}(s)=\frac{(d-2)!!}{s^2} +\sum_{m=0}^\infty \alpha_{n,m} H_{ij,m+1} s^{m},
\eeq
where
\beq  \label{Ak}
\alpha_{n,m} = \sum_{\ell=0}^{\min(n,m)} \frac{(-1)^{m-\ell}}{(m-\ell)!}
\omega_{n,\ell},
\eeq
\beq
H_{ij,m}= \int_0^\infty \dd r \, h_{ij}(r) r^{m}.
\eeq
\item[(iii)] Low-density expansion:
\beqa
\label{Grho}
G_{ij}(s)&=&\frac{\theta_{n+1}(\sigma_{ij}s)e^{-\sigma_{ij}s}}{s^2}+\rho \sum_{\ell=1}^\NN x_\ell \int_{\sigma_{ij}}^\infty \dd r\, r\theta_n(sr)\nn
&&\times
\Omega_{\sigma_{i\ell},\sigma_{j\ell}}(r)e^{-sr} +\cm{O}(\rho^2),
\eeqa
where
$\Omega_{a,b}(r)$ is the intersection volume of two hyperspheres of
radii ${a}$ and ${b}$ whose centers are separated by a distance $r\leq a+b$.
\end{itemize}
Conditions (i)--(iii) are consequences of the hard-core interaction and,
therefore, are directly related to the step-function structure of
$g_{ij}(r)$ [Eq.\ (\ref{ghard})]. Because $\alpha_{n,2q+1}=0$ for
$q=0,\ldots,n-1$ \cite{RS07}, condition (ii) guarantees through (\ref{hG}) that the
correlation functions remain bounded at the limit of zero wave-number,
\beq \label{hij0}
\widehat{h}_{ij}(0) < \infty,
\eeq
and  this in turn assures that the isothermal compressibility of the fluid
takes finite values \cite{RS11}.

\subsection{The PY approximation} \label{s.PY}

The PY approximation to a classical fluid is obtained from the
OZ equation (\ref{OZ}) supplemented by a particular closure relation between
$g_{ij}(r)$ and $c_{ij}(r)$.
For hard-hypersphere systems, the PY closure reduces to Eq.\ (\ref{ghard}) and
the assumption that the direct correlation function is short ranged, namely,
\beq \label{cPY}
c_{ij}(r)=0, \quad r>\sigma_{ij}.
\eeq
Therefore, if an approximate radial distribution function $g_{ij}(r)$ satisfies the hard-core condition \eqref{ghard} and its associated direct correlation function, as obtained from Eq.\ (\ref{OZ}), verifies  the condition (\ref{cPY}), it is necessarily a solution to the PY theory.

\subsection{The RFA} \label{s.RFA}

In contrast to the PY approach, the RFA method is based on the Laplace
functionals $G_{ij}(s)$ rather than on the OZ equation. Specifically,
the RFA  provides an analytical representation of $G_{ij}(s)$ that
complies with the consistency conditions (i)--(iii) \cite{RS11}.
This analytical approximation reads \cite{note_11_07}
\beq
\label{Grfa}
G_{ij}(s)=s^{d-2} e^{-\sigma_{ij}s}
\left[ \sm{L}(s)\cdot \sm{B}^{-1}(s)\right]_{ij},
\eeq
where $\sm{L}(s)$ and $\sm{B}(s)$ are $\NN\times\NN$ matrices with elements
\beq
\label{L}
L_{ij}(s)=\sum_{m=0}^{n+1} L_{ij}^{(m)} s^m,
\eeq
\beq \label{Bij}
B_{ij}(s) = s^d\delta_{ij} - \nu\rho_i e^{-\sigma_i s} L_{ij}(s)
+P_{ij}(s).
\eeq
In Eq.\ \eqref{Bij}, $\delta_{ij}$ is the Kronecker delta symbol,
$\rho_i=\rho x_i$ is the partial number density of species $i$, and
\beq
\label{Pij}
P_{ij}(s)=\nu\rho_i  \sum_{m=0}^{n+1} \left[ \sum_{\ell=0}^{d-m}
\frac{(-\sigma_is)^\ell}{\ell!} \right] L_{ij}^{(m)} s^m .
\eeq
Note that the matrices $\sm{L}(s)$ and $\sm{P}(s)$ have polynomial dependencies on $s$ of degrees $n+1=(d-1)/2$ and $2n+3=d$, respectively.

The coefficients $L_{ij}^{(m)}$, $m=0,1,...,n+1=(d-1)/2$, may depend on the fluid
density, the particle diameters, and the component abundances, but they are
independent of $s$. Those coefficients are determined from a set of $n+2$ algebraic
matrix equations stemming from the requirement of condition
(\ref{Gshort}).
The reader is urged to consult Ref.\ \cite{RS11} for further details.

Note that Eq.\ \eqref{Grfa} can be rewritten as
\beq
\label{Grfa2}
G_{ij}(s)=s^{d-2} e^{-\sigma_{ij}s}\frac{F_{ij}(s)}{\beta(s)},
\eeq
where
\beq \label{Fb}
\sm{F}(s) \equiv \sm{L}(s)\cdot \text{adj}[\sm{B}(s)],\quad \beta(s)\equiv |{\sm{B}}(s)|.
\eeq
Here, $\text{adj}(\sm{A})$ and $|\sm{A}|$ refer to the adjoint and the determinant, respectively, of a matrix $\sm{A}$.
Once the coefficients $L_{ij}^{(m)}$ are obtained as  functions of
$\rho$, $\{\sigma_i\}$, and  $\{x_i\}$, the functionals
$G_{ij}(s)$ are fully  determined from Eqs.\ (\ref{Grfa}) or (\ref{Grfa2}). Then, the total
correlation functions in $k$ space are obtained from Eq.\ (\ref{hG}), which
can be rewritten using (\ref{Grfa2}) as
\beq
\label{hrfa}
\widehat{h}_{ij}(k)=\nu \frac{R_{ij}(\ii k)}
{\beta(\ii k)\beta(-\ii k)},
\eeq
where
\beq \label{Rij}
R_{ij}(s)\equiv F_{ij}(s)\beta(-s)e^{-\sigma_{ij}s}
+F_{ij}(-s)\beta(s)e^{\sigma_{ij}s}.
\eeq
Finally, to obtain the direct correlation functions in the
configuration space, one makes use of Eq.\ (\ref{OZ}) and the inverse Fourier
transform \cite{RS11},
\beq \label{cr}
c_{ij}(r)=\frac{(2\pi)^{-(d+1)/2}}{r^{d-2}} \ii \int_{-\infty}^\infty
\dd k\, k \widehat{c}_{ij}(k)\theta_n(\ii kr) e^{-\ii kr}.
\eeq
It is worth noting that, as shown in Ref.\ \cite{RS11}, the Laplace
functionals obtained from (\ref{Grfa}) satisfy the asymptotic behaviors
given by Eqs.\ (\ref{Gsi})--(\ref{Grho}). In particular,  the physical hard-core
requirement given by Eq.\ (\ref{ghard}) is verified. Therefore, in order to prove the
equivalence between the PY and RFA approaches it will be sufficient to show
that the direct correlation functions calculated with Eqs.\ (\ref{OZ}),
(\ref{hrfa}), and (\ref{cr}) are short ranged,  as required by Eq.\ (\ref{cPY}).

Before closing this section, let us derive some properties for small $s$ that will be useful in Sec.\ \ref{s.analysis}.
By expanding the exponential in Eq.\ \eqref{Bij} and inserting Eq.\ \eqref{Pij}, one gets
\beq \label{Bij2}
B_{ij}(s) = s^d\delta_{ij} - \nu\rho_i  \sum_{m=0}^{n+1} \left[ \sum_{\ell=d-m+1}^{\infty}
\frac{(-\sigma_is)^\ell}{\ell!} \right] L_{ij}^{(m)} s^m.
\eeq
Therefore, $B_{ij}(s)=s^d[\delta_{ij}+\mathcal{O}(s)]$, so that the determinant is
\beq
\beta(s)=s^{\NN d}\left[1+\mathcal{O}(s)\right].
\label{beta(s)}
\eeq
Next, from Eqs.\ \eqref{Gshort} and \eqref{Grfa2} we get
\beq
e^{-\sigma_{ij}s}F_{ij}(s)=(d-2)!!\frac{\beta(s)}{s^d}\left[1+s^2\mathcal{F}_{ij}(s^2)+\mathcal{O}(s^d)\right],
\label{new1}
\eeq
where $\mathcal{F}_{ij}(s^2)$ is a polynomial of degree $n=(d-3)/2$ in $s^2$ whose explicit form will not be needed here.
In fact, the polynomial $\mathcal{F}_{ij}(s^2)$ cancels in Eq.\ \eqref{Rij}, resulting in $R_{ij}(s)=\mathcal{O}(s^{2\NN d})$.
{}From Eqs.\ \eqref{beta(s)} and \eqref{new1} one obtains
\beq
F_{ij}(s)=(d-2)!!s^{(\NN-1)d}\left[1+\mathcal{O}(s)\right].
\label{new1b}
\eeq

Particularizing to the binary case ($\NN=2$) and taking the determinant on both sides of Eq.\ \eqref{new1}, we get
\beq
e^{-2\sigma_{12} s}|\sm{F}(s)|=\left[\frac{\beta(s)}{s^d}\right]^2s^2\left[\mathcal{F}(s^2)+\mathcal{O}(s^{d-2})\right],
\label{|F(s)|}
\eeq
where $\mathcal{F}(s^2)$ is a polynomial of degree $n=(d-3)/2$ in $s^2$.
Combination of Eqs.\ \eqref{new1} and \eqref{|F(s)|} yields
\beqa
e^{(2\sigma_{12}-\sigma_{ij})s}|\sm{L}(-s)|F_{ij}(s)&=&\frac{\beta(-s)\beta(s)}{s^{3d-2}}\left[\mathcal{L}_{ij}(s^2)\right.\nn
&&\left.+\mathcal{O}(s^{d-2})\right],
\label{new2}
\eeqa
where  $\mathcal{L}_{ij}(s^2)$ is again a polynomial of degree $n=(d-3)/2$ and use has been made of the property
\beq
\label{detF}
|\sm{F}(s)|=\beta(s)|\sm{L}( s)|,
\eeq
which follows from  Eq.\ (\ref{Fb}).
Finally, Eq.\ \eqref{new2} implies
\beqa
e^{(2\sigma_{12}-\sigma_{ij})s}|\sm{L}(-s)|F_{ij}(s)
&+&e^{-(2\sigma_{12}-\sigma_{ij})s}|\sm{L}(s)|F_{ij}(-s)\nn
&=&\frac{\beta(-s)\beta(s)}{s^{3d-2}}\mathcal{O}(s^{d-2})\nn
&=&\mathcal{O}(s^{2d}),
\label{new3}
\eeqa
where in the last equality we have taken into account Eq.\ \eqref{beta(s)}.

\section{Binary mixtures} \label{s.binary}

Henceforth, we consider an additive system of two components ($\NN=2$).
From Eq.\ (\ref{OZ}), we have
\beq
\widehat{c}_{11}(k)=\frac{\widehat{h}_{11}(k)+\rho_2
 |\widehat{\sm{h}}(k)|}{\dt(k)},
\quad
\widehat{c}_{12}(k)=\frac{\widehat{h}_{12}(k)}{\dt(k)},
\label{c11k}
\eeq
with
\beq \label{dk}
\dt(k)\equiv 1 +\rho_1 \widehat{h}_{11}(k)+\rho_2 \widehat{h}_{22}(k)
+\rho_1\rho_2 |\widehat{\sm{h}}(k)|.
\eeq
The expressions for $\widehat{c}_{22}(k)$ and $\widehat{c}_{21}(k)$ are
obtained from those of $\widehat{c}_{11}(k)$ and $\widehat{c}_{12}(k)$
by exchanging the subscripts 1 and 2.

Equations \eqref{c11k} and \eqref{dk} are general and valid for any binary mixture. Now we particularize to the RFA. Using Eq.\ (\ref{hrfa}), one obtains
\beq \label{dD}
\dt(k)=\frac{\Q(\ii k)}{\beta(\ii k)\beta(-\ii k)},
\eeq
\beq \label{c11}
\widehat{c}_{11}(k)=\nu \frac{{R}_{11}(\ii k)
+\nu \rho_2 |\sm{R}(\ii k)|/[\beta(\ii k)\beta(-\ii k)]}{\Q(\ii k)},
\eeq
\beq \label{c12}
\widehat{c}_{12}(k)=\nu  \frac{{R}_{12}(\ii k)}
{\Q(\ii k)},
\eeq
where
\beqa
\Q(s)&\equiv&\beta(s)\beta(-s) +\nu[ \rho_1 R_{11}(s)+\rho_2 R_{22}(s)]\nn
&&
+\nu^2 \rho_1 \rho_2 \frac{|\sm{R}(s)|}{\beta(s)\beta(-s)}.
\eeqa
Explicit calculation using Eq.\ (\ref{Rij})  gives
\beq \label{Rbb}
 \frac{|\sm{R}(s)|}{ \beta(s)\beta(-s)} =\Phi(s)+\Phi(-s),
\eeq
\beq \label{Dbb}
\Q(s) = \beta(s)\beta(-s)+\Lambda(s)+\Lambda(-s),
\eeq
where
\beqa
\Phi(s) &\equiv& |\sm{L}(s)|\beta(-s) e^{-(\sigma_1+\sigma_2)s}- F_{12}(s)F_{21}(-s)
        \nn
        && + F_{11}(s)F_{22}(-s) e^{-(\sigma_1-\sigma_2)s},
        \label{Phi}
\eeqa
\beqa \label{lam}
\Lambda(s)&\equiv& \nu \beta(-s) [\rho_1 F_{11}(s)e^{-\sigma_1s}
           +\rho_2 F_{22}(s)e^{-\sigma_2s}] \nn
&&+\nu^2 \rho_1 \rho_2 \Phi(s).
\eeqn
In Eq.\ \eqref{Phi} we have used Eq.\ \eqref{detF}.

Thus far, we have not used in this section the explicit form of the matrix $\sm{B}(s)$, Eq.\ \eqref{Bij}.
Evaluation of Eq.\ (\ref{Fb}) for binary mixtures
yields
\beq \label{Fm}
\sm{F}(s)=
\begin{bmatrix}
 \wt{F}_{11}(s)-\nu \rho_2 |\sm{L}(s)|e^{-\sigma_2s} &
 \wt{F}_{12}(s) \\
 \wt{F}_{21}(s) &
 \wt{F}_{22}(s)-\nu \rho_1 |\sm{L}(s)|e^{-\sigma_1s}
\end{bmatrix},
\eeq
where
\beq
\wt{F}_{11}(s) \equiv  L_{11}(s)[s^d+P_{22}(s)]-L_{12}(s)P_{21}(s),
\label{F11t}
\eeq
\beq
\wt{F}_{12}(s) \equiv  L_{12}(s)[s^d+P_{11}(s)]-L_{11}(s)P_{12}(s).
\label{F12t}
\eeq
Of course, $\wt{F}_{22}(s)$ and $\wt{F}_{21}(s)$ are obtained from Eqs.\ (\ref{F11t}) and \eqref{F12t} by
exchanging the subscripts 1 and 2. In addition, the determinant $\beta(s)$
of the matrix $\sm{B}(s)$ can be expressed as
\beqa
\label{be}
\beta(s)&=& \wt{\beta}(s) +\nu^2\rho_1\rho_2 |\sm{L}(s)|
e^{-(\sigma_1+\sigma_2)s}\nn
&&
 -\nu\rho_1 \wt{F}_{11}(s)e^{-\sigma_1s}
-\nu\rho_2 \wt{F}_{22}(s)e^{-\sigma_2s},
\eeqa
with
\beq
\wt{\beta}(s)\equiv s^{2d}+s^d[P_{11}(s)+P_{22}(s)]+|\sm{P}(s)|.
\eeq

We can observe that $\wt{F}_{ij}(s)$ and $\wt{\beta}(s)$ are polynomials in $s$ of degrees $3n+4=(3d-1)/2$ and $2(2n+3)=2d$, respectively. In contrast, the functions $F_{11}(s)$, $F_{22}(s)$, $\beta(s)$, $\Phi(s)$, and $\Lambda(s)$ contain exponential terms. The key point, however, is that those exponential terms compensate exactly in the function $\Q(s)$. Using Eqs.\ \eqref{Phi}, \eqref{lam}, \eqref{Fm}, and \eqref{be} in Eq.\ \eqref{Dbb} one finds, after some algebra,
\beqa
\label{Dbb2}
\Q(s)&=& \wt{\beta}(s)\wt{\beta}(-s)
   + \nu^4 \rho_1^2 \rho_2^2 |\sm{L}(s)||\sm{L}(-s)|\nn
&& -\nu^2\sum_{i,j=1}^2\rho_i\rho_j\wt{F}_{ij}(s) \wt{F}_{ji}(-s).
\eeqa
Thus, $\Q(s)$ is an even polynomial of degree $4(2n+3)=4d$.

The direct correlation functions in  Fourier space are given by Eqs.\ (\ref{c11}) and \eqref{c12}.
{}From Eqs.\ \eqref{Rij}, \eqref{Rbb}, \eqref{Phi}, \eqref{Fm}, and \eqref{be} it is possible to get
\beqa
\label{cij}
\widehat{c}_{ij}(k) &=& \frac{\nu}{\Q(\ii k)}
   \left[\widetilde{\mathcal{P}}_{ij}(\ii k)e^{\ii\sigma_{ij}k}
   +\widetilde{\mathcal{P}}_{ij}(-\ii k)e^{-\ii\sigma_{ij}k}\right.\nn
   &&
   \left.+\widetilde{\mathcal{Q}}_{ij}(\ii k)e^{\ii(\sigma_i-\sigma_{j})k/2}
   +\widetilde{\mathcal{Q}}_{ij}(-\ii k)e^{\ii(\sigma_j-\sigma_{i})k/2}\right],\nn
\eeqa
where $\widetilde{\mathcal{P}}_{ij}(s)$ and $\widetilde{\mathcal{Q}}_{ij}(s)$ are polynomials of degrees $7n+10=(7d-1)/2$ and $2(3n+4)=3d-1$, respectively, given
by
\beq \label{P11}
\widetilde{\mathcal{P}}_{11}(s)= \wt{F}_{11}(-s)\wt{\beta}(s)
 -\nu^2\rho_2^2|\sm{L}(-s)|\wt{F}_{22}(s),
\eeq
\beq
\label{P12}
\widetilde{\mathcal{P}}_{12}(s)= \wt{F}_{12}(-s)\wt{\beta}(s)
 +\nu^2\rho_1\rho_2|\sm{L}(-s)|\wt{F}_{12}(s),
\eeq
\beqa
\label{Q11}
\widetilde{\mathcal{Q}}_{11}(s)&=& -\nu\left[\rho_1 \wt{F}_{11}(-s)\wt{F}_{11}(s)
 +\rho_2 \wt{F}_{12}(-s)\wt{F}_{21}(s)\right]\nn
 &&
 +\nu^3\rho_1\rho_2^2 |\sm{L}(s)||\sm{L}(-s)|,
\eeqa
\beq
\label{Q12}
\widetilde{\mathcal{Q}}_{12}(s)= -\nu\left[\rho_1 \wt{F}_{11}(-s)\wt{F}_{12}(s)
 +\rho_2 \wt{F}_{12}(-s)\wt{F}_{22}(s)\right].
\eeq
Again,
 $\widetilde{\mathcal{P}}_{22}(s)$, $\widetilde{\mathcal{P}}_{21}(s)$,
$\widetilde{\mathcal{Q}}_{22}(s)$, and $\widetilde{\mathcal{Q}}_{21}(s)$ are obtained by exchanging the subscripts 1 and 2.

\section{Analysis} \label{s.analysis}
In the derivation of the results in Sec.\ \ref{s.binary} we have not needed to use either Eq.\ \eqref{Pij} or the conditions that the coefficients $L_{ij}^{(m)}$ must satisfy [which are summarized by Eq.\ \eqref{new1}]. In fact, Eq.\ \eqref{cij} alone is not sufficient to prove the PY condition \eqref{cPY}.

As said before, the RFA method guarantees that $\widehat{h}_{ij}(k)$ takes
finite values at the limit of zero wave-number, Eq.\ (\ref{hij0}).
Thus, from Eq.\ (\ref{dk}), $\dt(k)$ remains bounded at $k=0$ and, according to Eq.\
(\ref{dD}), we have
\beq \label{d0}
\lim_{s\rightarrow 0} \frac{\Q(s)}{\beta(s)\beta(-s)}
=\dt(0) < \infty.
\eeq
Taking into account that, in the binary case ($\NN=2$), $\beta(s)=s^{2d}$ for small $s$ [see  Eq.\ \eqref{beta(s)}], we
have $\Q(s)=\cm{O}(s^{4d})$. Since, according to Eq.\ \eqref{Dbb2},  $\Q(s)$ is a polynomial of degree $4d$, we conclude  that  the $\Q(s)$ is just a pure power law, i.e.,
\beq
\Q(s) =\dt(0) s^{4d}.
\label{new4}
\eeq
This is the crucial result allowing one to prove Eq.\ \eqref{cPY} from Eq.\ \eqref{cij}. Before proceeding to the proof, let us first simplify Eq.\ \eqref{cij} a little more.

As shown by Eqs.\ \eqref{P11}--\eqref{Q12}, $\widetilde{\mathcal{P}}_{ij}(s)$ and $\widetilde{\mathcal{Q}}_{ij}(s)$ are polynomials of degrees $7n+10=(7d-1)/2$ and $2(3n+4)=3d-1$, respectively. Going back to the quantities $F_{ij}(s)$ and $\beta(s)$, Eqs.\ \eqref{P11}--\eqref{Q12} can be rewritten as
\beqa
\widetilde{\mathcal{P}}_{ij}(s)&=&\widetilde{F}_{ij}(-s)\beta(s)+\nu F_{ij}(-s)\sum_{k=1}^2\rho_k e^{-\sigma_k s}F_{kk}(s) \nn&&+\nu^2\rho_1\rho_2 e^{-\sigma_{ij} s}\left[e^{(2\sigma_{12}-\sigma_{ij}) s}|\sm{L}(-s)|F_{ij}(s)\right.\nn&&\left.+e^{-(2\sigma_{12}-\sigma_{ij}) s}|\sm{L}(s)|F_{ij}(-s)\right],
\label{cPij}
\eeqa
\beqa
\widetilde{\mathcal{Q}}_{ij}(s)&=&-\nu \sum_{k=1}^2\rho_k F_{ik}(-s)F_{kj}(s) \nn&&-\nu^2\rho_1\rho_2 e^{-(\sigma_{i}-\sigma_j) s/2}\left[e^{(2\sigma_{12}-\sigma_{ij}) s}|\sm{L}(-s)|F_{ij}(s)\right.\nn&&\left.+e^{-(2\sigma_{12}-\sigma_{ij}) s}|\sm{L}(s)|F_{ij}(-s)\right].
\label{cQij}
\eeqa

While Eqs.\ \eqref{cPij} and \eqref{cQij} conceal the polynomial character of $\widetilde{\mathcal{P}}_{ij}(s)$ and $\widetilde{\mathcal{Q}}_{ij}(s)$, they show, with the help of Eqs.\ \eqref{beta(s)}, \eqref{new1b}, and \eqref{new3}, that
\beq
\widetilde{\mathcal{P}}_{ij}(s)=\mathcal{O}(s^{2d}),\quad \widetilde{\mathcal{Q}}_{ij}(s)=\mathcal{O}(s^{2d}).
\eeq
This result, along with Eq.\ \eqref{new4}, allows us to rewrite Eq.\ \eqref{cij} as
\beqa
\label{cijbis}
\widehat{c}_{ij}(k) &=& \frac{1}{k^{2d}}
   \left[\mathcal{P}_{ij}(\ii k)e^{\ii\sigma_{ij}k}
   +\mathcal{P}_{ij}(-\ii k)e^{-\ii\sigma_{ij}k}\right.\nn
   &&
   \left.+\mathcal{Q}_{ij}(\ii k)e^{\ii(\sigma_i-\sigma_{j})k/2}
   +\mathcal{Q}_{ij}(-\ii k)e^{\ii(\sigma_j-\sigma_{i})k/2}\right],\nn
\eeqa
where
\beq
\mathcal{P}_{ij}(s)=-\frac{\nu}{\dt(0)}s^{-2d}\widetilde{\mathcal{P}}_{ij}(s)
\label{new5}
\eeq
is a polynomial of degree $3n+4=3(d-1)/2$ and
\beq
\mathcal{Q}_{ij}(s)=-\frac{\nu}{\dt(0)}s^{-2d}\widetilde{\mathcal{Q}}_{ij}(s)
\label{new6}
\eeq
is a polynomial of degree $2(n+1)=d-1$.

Equation \eqref{cijbis} is the main result of this paper. It provides the  functional structure of  the Fourier transform $\widehat{c}_{ij}(k)$ in the RFA
approach.  The direct correlations function in the configuration space $c_{ij}(r)$ can be
calculated by application of the residue theorem  combining Eqs.\ (\ref{cr}) and (\ref{cijbis}).
Since $\widehat{c}_{ij}(0)=\mbox{finite}$, the integrand in Eq.\
(\ref{cr}) is regular along the real axis and so we
can distort the integration path in the complex $k$ plane by going around the point $k=0$ from
below. Next,
the integral in Eq.\ (\ref{cr}) decomposes into four contributions with
integrands headed by $e^{-\ii k(r-\sigma_{ij})}$, $e^{-\ii k(r+\sigma_{ij})}$,
$e^{-\ii k[r+(\sigma_j-\sigma_i)/2]}$, and
$e^{-\ii k[r+(\sigma_i-\sigma_j)/2]}$, respectively, with each integrand having a single pole at $k=0$ of order ${2d-1}$.
If $0<r<\sigma_{ij}$, the first integral must be closed with an
upper half circle of infinite radius, and the residue theorem yields a nonzero value. If
$0<r<|\sigma_j-\sigma_i|/2$, an additional nonzero contribution results from the third
or fourth integral, depending on whether $\sigma_i>\sigma_j$ or $\sigma_j>\sigma_i$,
respectively.
On the other hand, if $r>\sigma_{ij}$, we
must close the path  with a lower half circle. As a consequence, the
four contributions vanish, and so the RFA method yields Eq.\ \eqref{cPY}.

This completes the proof on the equivalence between the PY and RFA
solutions for binary mixtures of hard hyperspheres at odd dimensional space.

\section{Concluding remarks} \label{s.conclusions}

PY and RFA theories are, in principle, alternative methods for calculating thermodynamic and
structural functions of HS systems.
The PY theory consists of the OZ relation \eqref{OZ} supplemented with the hard-core condition \eqref{ghard} and the genuine PY closure \eqref{cPY}.
In the RFA approach, however, one proposes a specific form, Eqs.\ \eqref{Grfa}--\eqref{Pij}, for the $s$-dependence of the Laplace functional $G_{ij}(s)$ defined by Eq.\ \eqref{Gij}. This specific form includes $n+2=(d+1)/2$ coefficients, $L_{ij}^{(m)}$, which are determined, in consistency with Eq.\ \eqref{Gshort},  by requiring the independent term in the Taylor series expansion of $s^2 G_{ij}(s)$ to be $(d-2)!!$ and all the coefficients of $s^{2q+1}$ with $q=0,\ldots,n$ to vanish.

In this paper we have shown by a direct verification that both methods are fully equivalent
for binary mixtures of additive hard hyperspheres at odd dimensions. The proof
is based on the analysis of the Fourier transform $\widehat{c}_{ij}(k)$ of the direct correlation function $c_{ij}(r)$ and proceeds along two main stages. In the first stage, use of Eqs.\ \eqref{Grfa}--\eqref{Pij} has allowed us to derive Eq.\ \eqref{cij}, where $\widehat{\mathcal{P}}_{ij}(s)$, $\widehat{\mathcal{Q}}_{ij}(s)$, and $\Q(s)$ are polynomials. This result applies regardless of the values of the $n+2$ coefficients $L_{ij}^{(m)}$. In the second stage, we have proved that enforcement of Eq.\ \eqref{Gshort} implies Eqs.\ \eqref{new4}, \eqref{new5}, and \eqref{new6}, so that Eq.\ \eqref{cij} simplifies further to Eq.\ \eqref{cijbis}. Application of the residue theorem then yields Eq.\ \eqref{cPY}, which completes the proof.
An interesting feature of the proof is that the explicit expressions for the coefficients $L_{ij}^{(m)}$  as functions of the physical parameters of the fluid
(density, mole fractions, and particle diameters) are not needed.

Although the proof presented in this paper has been restricted to the binary case ($\NN=2$), we conjecture that the structure of Eq.\ \eqref{cijbis}, and hence the validity of Eq.\ \eqref{cPY}, remains valid  for any number $\NN$ of components in the framework of the RFA.

It is worth mentioning that, as done for three-dimensional mixtures \cite{YSH98} and for $d$-dimensional one-component systems \cite{RS07}, the RFA scheme can be extended beyond the PY level by adding extra terms ${L}_{ij}^{(n+2)}$ in Eqs.\ \eqref{L} and \eqref{Pij}, and replacing $\delta_{ij}$ by $(1+u s)\delta_{ij}$ in Eq.\ \eqref{Bij}. The $\NN^2+1$ free  parameters ${L}_{ij}^{(n+2)}$ and  $u$  can be fixed by imposing given expressions for the contact values $g_{ij}(\sigma_{ij}^+)$  and the thermodynamically consistent isothermal compressibility.

\acknowledgments
{The work of {R.D.R.} has been supported by the Consejo Nacional de Investigaciones Cient\'ificas y
T\'ecnicas (CONICET, Argentina) through Grant No.\
PIP 112-200801-01474.  A.S. acknowledges support from the Ministerio de Ciencia e Innovaci\'on (Spain) through Grant No.\ FIS2010-16587 and  the Junta de Extremadura (Spain) through Grant No.\ GR10158, partially financed by Fondo Europeo de Desarrollo Regional (FEDER) funds.}

\bibliographystyle{apsrev}
\bibliography{D:/bib_files/liquid}

\begin{thebibliography}{11}
\bibitem[{not()}]{note_11_07}
\bibinfo{note}{For convenience, the matrix $\sm{B}(s)$ introduced here differs
  from that in Ref.\ \protect\cite{RS11} by a multiplicative factor $s^d$.}
\end{thebibliography}
\end{document}